\documentclass[10pt,twocolumn,letterpaper]{article}

\usepackage[cvpr]{cvpr}      

\usepackage{amsmath}
\usepackage{amssymb}
\usepackage{booktabs}
\usepackage{balance}

\usepackage{times,graphicx,amsmath,amssymb,booktabs,tabulary,multirow,overpic,xcolor}
\usepackage{colortbl}
\usepackage{makecell}
\usepackage{float}
\usepackage{setspace}
\definecolor{sh_gray}{rgb}{0.84,0.84,0.84}
\definecolor{sh_gray2}{rgb}{1,0.89,0.75}
\definecolor{color3}{rgb}{0.95,0.95,0.95}
\definecolor{color4}{rgb}{0.96,0.96,0.86}
\definecolor{color5}{rgb}{0.90,0.90,0.90}

\usepackage[pagebackref,breaklinks,colorlinks]{hyperref}

\usepackage[capitalize]{cleveref}
\usepackage{multirow}
\crefname{section}{Sec.}{Secs.}
\Crefname{section}{Section}{Sections}
\Crefname{table}{Table}{Tables}
\crefname{table}{Tab.}{Tabs.}
\usepackage{algorithm}
\usepackage{algorithmic}
\usepackage{enumitem}
\def\cvprPaperID{2182} 
\def\confName{CVPR}
\def\confYear{2022}

\begin{document}
\begin{spacing}{0.965}
\title{HDNet: High-resolution Dual-domain \\ Learning  for Spectral Compressive Imaging}
\author{Xiaowan Hu$^{*,1,2}$, Yuanhao Cai\thanks{Equal Contribution. $\dagger$ Corresponding Author.}$^{~ ,1,2}$, Jing Lin$^{1,2}$, Haoqian Wang$^{1,2,\dagger}$,\\ Xin Yuan$^{3}$, Yulun Zhang$^{4}$, Radu Timofte$^{4}$, and Luc Van Gool$^{4}$\\
	$^1$ Shenzhen International Graduate School, Tsinghua University\\
	$^2$ Shenzhen Institute of Future Media Technology,
	$^3$ Westlake University, $^4$  ETH Zurich
}

\maketitle

\begin{abstract}
The rapid development of deep learning provides a better solution for the end-to-end reconstruction of hyperspectral image (HSI). However, existing learning-based methods have two major defects. Firstly, networks with self-attention usually sacrifice internal resolution to balance model performance against complexity, losing fine-grained high-resolution (HR) features. Secondly, even if the optimization focusing on spatial-spectral domain learning (SDL) converges to the ideal solution, there is still a significant visual difference between the reconstructed HSI and the truth. Therefore, we propose a high-resolution dual-domain learning network (HDNet) for HSI reconstruction. On the one hand, the proposed HR spatial-spectral attention module with its efficient feature fusion provides continuous and fine pixel-level features. On the other hand, frequency domain learning (FDL) is introduced for HSI reconstruction to narrow the frequency domain discrepancy. Dynamic FDL supervision forces the model to reconstruct fine-grained frequencies and compensate for excessive smoothing and distortion caused by pixel-level losses. The HR pixel-level attention and frequency-level refinement in our HDNet mutually promote HSI perceptual quality. Extensive quantitative and qualitative evaluation experiments show that our method achieves SOTA performance on simulated and real HSI datasets. Code and models are released at  \url{https://github.com/caiyuanhao1998/MST}

\end{abstract}

\vspace{-4mm}
\section{Introduction}
\vspace{-2mm}

Hyperspectral images (HSIs) with more spectral bands can capture richer scene information and fixed wavelength imaging characteristics, which are used in image classification~\cite{intro_reg}, object detection~\cite{intro_od}, and tracking~\cite{ot_2,ot_1} widely.

Imaging systems with single 1D or 2D sensors take a long time to scan a scene for HSIs. They are not suitable for capturing dynamic scenes. Recently, the coded aperture snapshot spectral imaging (CASSI) system~\cite{sci_1,sparse_3,sci_3} can capture 3D HSI cubes at a real-time rate. CASSI exploits a coded aperture to modulate the HSI signal and compress it into a 2D measurement. Nonetheless, a core problem of the CASSI system is to recover the reliable and fine underlying 3D HSI signal from the 2D compressed images.
\begin{figure}[t]
 \flushleft
\hspace{-1.5mm}
\includegraphics[width=84mm]{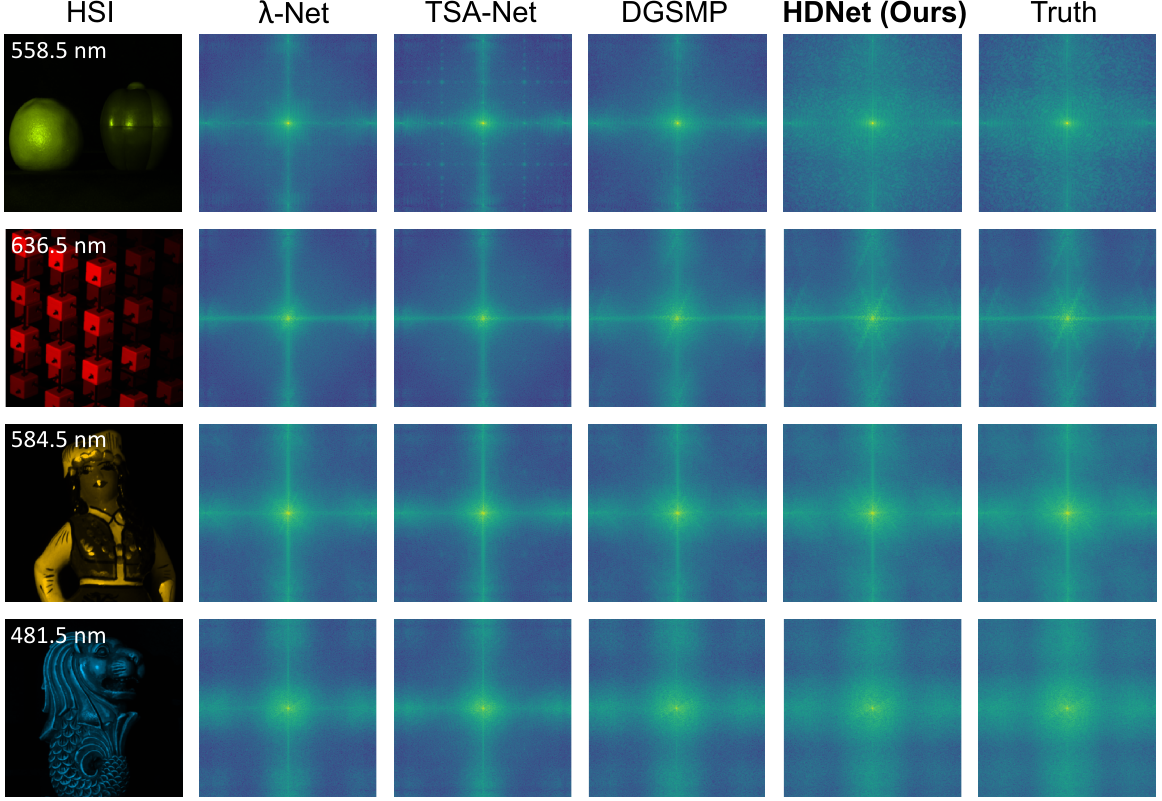}
\vspace{-7mm}
\caption{Frequency spectra visualization of different methods.}
\label{fig:intro}
\vspace{-6mm}
\end{figure}

Traditional methods mainly regularize the reconstruction based on hand-crafted priors that delineate the structure of HSI. But manually adjusted parameters result in poor generalization. Inspired by the success of deep learning in image restoration, researchers began using  convolutional neural networks (CNN) for HSI reconstruction. Some methods~\cite{lambda,desci,tsa_net,dnu} focus on self-attention learning in the spatial HSI domain, but they usually sacrifice feature resolution to cut down the computational complexity of non-local attention maps~\cite{lambda,tsa_net}. These operations inevitably damage the spectral auto-correlation and information continuity. Inspired by the extensive exploration of pixel-level attention in high-level visual tasks~\cite{cheng2020panoptic,liu2021polarized}, we find that it is critical to elaborate high-resolution (HR) and fine-grained spectral-spatial attention for HSIs. However, although finer attention is undoubtedly beneficial for reconstructing HSI with rich spectral bands, capturing pixel-level perception for HSI with 28 spectral channels is far more challenging than 3-channel RGB images. It requires an optimal trade-off between model performance and resource costs.

Besides, existing learning-based methods~\cite{hssp,lambda,tsa_net,gsm} for HSI reconstruction mainly focus on {spatial-spectral domain learning (SDL), where the spectral representations are sparsely presented in the frequency domain.} The equal treatment of each frequency may result in sub-optimal mode efficiency. Some works show that due to the inherent biases of CNNs~\cite{tancik2020fourier,rahimi2007random,rahaman2019spectral,mildenhall2020nerf}, models tend to preferentially fit low-frequency components that are easy to synthesize while losing high-frequency components. We visualize the spectra of reconstructed HSIs in the frequency domain in Fig.~\ref{fig:intro}. We can see that even if previous methods based on SDL converge to the ideal solution, there is still an obvious frequency domain discrepancy between these reconstructed HSIs and ground truth. TSA-Net~\cite{tsa_net} loses high-frequency information and has observable checkerboard artifacts. DGSMP~\cite{gsm} deviates to a limited frequency area. The focal frequency loss~\cite{jiang2021focal} is proven to be effective in synthesizing fine frequency components, but its potential to narrow the frequency domain gap in HSI reconstruction still remains under-explored. We find that each frequency in the spectra is the statistical sum across all pixels in the HSI, so the frequency-level supervision can offer a new solution for global optimization. Experiments show that frequency domain learning (FDL) can compensate for excessive smoothing and distortion caused by pixel-level SDL.
 
Motivated by these meaningful findings, we propose a high-resolution dual-domain learning network, dubbed HDNet. Dual-domain supervision exhausts the model representation capacity within its spatial-spectral domain and frequency domain. On the one hand, in the spatial-spectral domain, we split the feature as HR spectral attention and HR spatial attention, and connect them in an efficient feature fusion (EFF) manner. The proposed fine-grained pixel-level attention avoids dimensionality collapse for high internal resolution. On the other hand, we use the Discrete Fourier Transform (DFT) to supervise the frequency distance between truth and reconstructed HSIs adaptively. Dynamic weighting mechanism makes the model concentrate high frequencies that are difficult to synthesize. Frequency spectra reconstructed by HDNet are the closest to the truth in Fig.~\ref{fig:intro}, which shows our superiority in narrowing the frequency difference between HSIs. The HR pixel-level attention in SDL and frequency-level refinement in FDL mutually promote common prosperity and ameliorate image quality further. 
Specific contributions of this paper are:
\begin{itemize}[leftmargin=*]
\setlength{\itemsep}{2pt}
\setlength{\parsep}{2pt}
\setlength{\parskip}{2pt}
	\item Dynamic frequency-level supervision that can narrow the frequency domain discrepancy is first used to improve the perceptual quality of HSIs. The proposed FDL forces the model to restore high and hard frequencies adaptively.
	\item We design the HR pixel-level attention in SDL for higher internal feature resolution, which further assists frequency alignment in FDL. Complementary dual-domain learning mechanism ameliorate HSI quality mutually.
	\item Our method achieves state-of-the-art (SOTA) performance in quantitative evaluation and visual comparison. Extensive experiments prove the superiority of HDNet.
\end{itemize}

\section{Related Work}
\subsection{HSI Reconstruction}
\vspace{-1.5mm}
Traditional methods~\cite{sparse_1,sparse_2,sparse_3,desci,non_local_1,non_local_2,gap_tv,tra_rela_1} mainly recover the 3D HSI cube from the 2D compressive measurement based on hand-crafted priors. Nonetheless, these model-based methods suffer from poor generalization ability. Inspired by the success of deep learning, researchers have started using deep CNN for HSI reconstruction~\cite{lambda,hssp,dnu,tsa_net,gsm}. GAP-Net~\cite{gapnet} presents a deep unfolding method and utilizes a pre-trained denoiser for HSI restoration. $\lambda$-Net~\cite{lambda} and TSA-Net~\cite{tsa_net} explores the self-attention for spatial features. The DGSMP~\cite{gsm} uses the deep Gaussian scale mixture prior for promising HSI reconstruction. However, current learning-based methods mainly focus on the spatial-spectral domain, where the frequency domain learning for HSI reconstruction remains under-investigated.
\vspace{-2.5mm}
\subsection{Self-Attention Mechanism}
\vspace{-1.5mm}
The self-attention mechanism~\cite{vaswani2017attention,shaw2018self} is widely used to capture long-range interactions. Many attention module and its variants used for natural images have shown great potential~\cite{wang2018non,hu2018squeeze,hu2018gather,shen2021efficient,woo2018cbam}. The $\lambda$-Net~\cite{lambda} first explored the feature auto-correlation in HSI restoration. Then ~\cite{mei2019spectral} uses the bi-directional network to model spectral correlation. The TSA-Net~\cite{tsa_net} calculates the spatial attention map and the spectral attention map separately. Wang \emph{et al.}~\cite{dnu} utilize the local and non-local correlation between spectral images. However, most existing networks sacrifice the internal resolution of attention to speed up calculations, which inevitably degrades performance. There are some pixel-level attention modules designed for high-level tasks~\cite{zhou2019objects,cheng2020panoptic,liu2021polarized} further enhances the model's representation ability. Therefore, the exploration of pixel-level HR attention for HSI reconstruction can provide a targeted solution for boosting performance.
\begin{figure*}[ht] 
\flushleft  
\includegraphics[width=176mm]{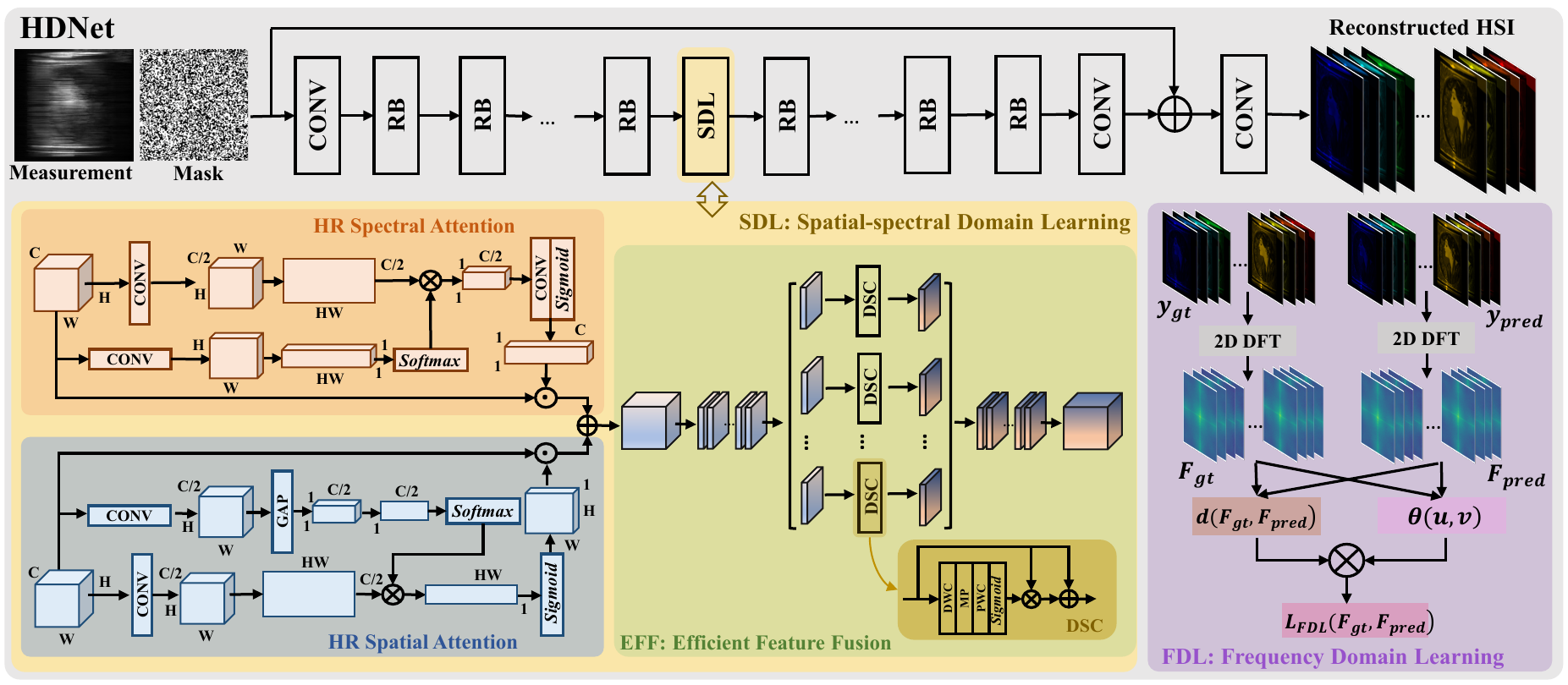}
\vspace{-7mm}
\caption{The architecture of HDNet. Spatial-spectral domain learning (SDL) includes HR spectral attention, HR spatial attention, and efficient feature fusion (EFF). In frequency domain learning (FDL), the 2D Discrete Fourier Transform (DFT) is used to obtain the HSI frequency spectrum. The adaptive weight $\theta(u,v)$ of each frequency coordinate $(u,v)$ is dynamically determined by the frequency distance.}
\label{overall}
\vspace{-4mm}
\end{figure*}
\vspace{-2mm}
\subsection{Image Frequency Spectrum Analysis}
\vspace{-1.5mm}
Frequency spectrum analysis describes signal frequency characteristics~\cite{tancik2020fourier,rahimi2007random}. The F-Principle~\cite{xu2019training} prove that deep learning networks tend to prefer low frequencies to fit the objective, which will result in the frequency domain gap~\cite{xu2019frequency,zhang2019explicitizing}. Recent studies~\cite{wang2020cnn,zhang2019detecting,huang2020fakeretouch} indicate that the periodic pattern shown in the frequency spectrum may be consistent with the artifacts in the spatial domain. Therefore, some works try to reduce the visual difference by narrowing the frequency domain gap between the input and output. \cite{fritsche2019frequency} treats low-frequency and high-frequency images differently during training. DASR~\cite{wei2021unsupervised} uses domain-gap aware training and domain-distance weighted supervision to solve the domain deviation in super-resolution. Jiang \emph{et al.}~\cite{jiang2021focal} proved that focusing on difficult frequencies can improve the reconstruction quality. In HSI reconstruction, the model over-fitting at low frequencies brings smooth textures and blurry structures. So exploring adaptive constraints on specific frequencies is essential for the refined reconstruction.
\end{spacing}

\vspace{-1.5mm}
\section{The Proposed Method}
\vspace{-1mm}
\subsection{Overall Architecture}
\vspace{-1.5mm}
The overall network architecture and internal module details of our proposed HDNet are shown in Fig.~\ref{overall}. We choose ResNet~\cite{he2016deep} as the baseline to build the proposed HDNet for a lightweight model, which is convenient to show the superiority of our designed plug-and-play components. 

In CASSI, the mask $\textbf{m} \in \mathbb{R}^{H\times W}$ is used to modulate the HSI signals. Then the modulated HSIs are shifted in the dispersion process. Thus, we shift back the measurement $\textbf{y} \in \mathbb{R}^{H\times (W+d(N_{\beta} - 1))}$, where $d$ denotes the shifting interval and $N_{\beta}$ denotes the number of wavelengths in HSIs. Then we formulate the dispersion process as follow:
\setlength{\abovedisplayskip}{4pt}
\setlength{\belowdisplayskip}{4pt}
\begin{align}
\textbf{y}' (x,y,n_{\beta})=\textbf{y} (x,y-d(\beta_{n} - \beta_{r})),
\label{input_1}
\end{align}
where $\textbf{y}'$ represents the multi-channel shifted version of the measurement, $n_{\beta} \in \{1,...,N_{\beta}\}$ indices the spectral channel, $\beta{r}$ is supposed to the reference wavelength, and $d(\beta_{n} - \beta_{r})$ indicates the shifted distance for the $n_{\beta}$-th channel. Then we use the mask to modulate $\textbf{y}'$ for the input:
\setlength{\abovedisplayskip}{4pt}
\setlength{\belowdisplayskip}{4pt}
\begin{align}
\textbf{x}_{in} = \textbf{y}' \odot \textbf{m},
\label{input_2}
\end{align}
where $\odot$ denotes the element-wise product. Then we define the $3 \times 3$ convolution layer as ${F}_{conv}^{3\times 3}(\cdot)$ to extract shallow features, and the corresponding features $\textbf{x}_0$ is defined as:
\setlength{\abovedisplayskip}{4pt}
\setlength{\belowdisplayskip}{4pt}
\begin{align}
\textbf{x}_0={F}_{conv}^{3\times 3}(\textbf{x}_{in}).
\label{fs}
\end{align}

To prove the effectiveness and efficiency of the spatial-spectral domain learning module, we only insert an SDL block in the middle of the stacked residual blocks (RBs). We define the number of RB stacked before and after SDL as $l$ and $g$ respectively, and we process the input as follow:

\setlength{\abovedisplayskip}{2pt}
\setlength{\belowdisplayskip}{4pt}
\begin{equation}
 \begin{split}
\textbf{x}_t &= {F}^l_{RB}(\textbf{x}_{l-1})={F}^l_{RB}({F}^{l-1}_{RB}(\cdot \cdot \cdot({F}^1_{RB}(\textbf{x}_0))\cdot \cdot \cdot)),\\
\Hat{\textbf{x}}_{f} &= {F}_{SDL}(\textbf{x}_{t}),\\
\Hat{\textbf{x}}_t &= {F}^g_{RB}(\textbf{x}_{g-1})={F}^g_{RB}({F}^{g-1}_{RB}(\cdot \cdot \cdot({F}^1_{RB}(\Hat{\textbf{x}}_{f}))\cdot \cdot \cdot)),
 \end{split}
\end{equation}
where the ${F}_{RB}(\cdot)$ and ${F}_{SDL}(\cdot)$ correspond to the RB and SDL module functions. The SDL module includes HR spectral attention, HR spatial attention, and efficient feature fusion (EFF). The internal implementation details of these modules will be described in Section~\ref{sec:sdl}. To maintain the high internal resolution of the features extracted from the stacked RBs, we use feature reshape and matrix multiplication operations instead of feature downsampling and channel sharp narrowing operations, and the grouped split-and-merge structure is designed for efficient feature fusion. 

\vspace{-3.5mm}
Global skip connection combines the shallow features $\textbf{x}_0$ with the deep features to further increase the model stability and information flow. Through the channel adjustment of the convolutional layers, we get the reconstructed HSI as:
\setlength{\abovedisplayskip}{4pt}
\setlength{\belowdisplayskip}{4pt}
\begin{align}
y_{pred}= {F}_{conv}^{3\times 3}({F}_{conv}^{3\times 3}(\Hat{\textbf{x}}_t) + \textbf{x}_0).
\label{xf}
\end{align}
As shown in Fig.~\ref{overall}, the predicted HSI $y_{pred}$ is supervised by a dual-domain learning mechanism. The SDL module is designed to be plug-and-play, and the FDL mechanism is used for loss optimization, which constrains the frequency distance between the reconstructed HSI and the truth adaptively. Dynamic weighting mechanism makes the model focus on hard frequencies reconstruction that is easily ignored by SDL. The HR pixel-level attention in SDL and the frequency-level refinement in FDL to achieve complementary learning, which further ameliorates HSI quality. In the following, we will introduce these two domains in detail.

\vspace{-1mm}
\subsection{Spatial-Spectral Domain Learning}
\label{sec:sdl}
\vspace{-1.5mm}
We extract the HR spatial-spectral attention cross spectral and spatial direction respectively, and perform efficient feature fusion (EFF) for more efficient representation.

\noindent\textbf{HR Spectral Attention.} We use two convolution layers to obtain \textit{query} vector defined as $f_c^{q}$ with the full spatial resolution and \textit{key} vector defined as $f_c^{k}$ with the half channel resolution. Then the \textit{query} vector does attention remapping to the \textit{key} vector for the \textit{value} vector $f_c^{v}$, and its spectral dimension remains $
C/2$, which avoids excessive loss of continuity. The the input $\textbf{x}_t\in\mathbb{R}^{\it H \times \it W \times \it C}$ is processed as:
\setlength{\abovedisplayskip}{4pt}
\setlength{\belowdisplayskip}{4pt}
\begin{equation}
 \begin{split}
 f_c^{q} &= {F}_{conv}^{1\times 1}(\textbf{x}_t)\in \mathbb{R}^{\it 1 \times \it H \times \it W },\\
f_c^{k} &=  {F}_{conv}^{1\times 1}(\textbf{x}_t)\in \mathbb{R}^{\textit{C / }2 \times \it H \times \it W},\\
f_c^{v} &= {F}_R(f_c^{k})\otimes Softmax[{F}_R(f_c^{q})]\in \mathbb{R}^{\textit{C / }2 \times \it1 \times \it1},
 \end{split}
\end{equation}
where ${F}_{conv}^{1\times 1}(\cdot)$ is the $1\times 1$ convolution function and the reshape function ${F}_R(\cdot)$ is used to facilitate size matching. $\otimes$ means matrix multiplication operation. In Fig.~\ref{overall}, after channel adjustment and \textit{Sigmoid} activation, the weight factor of each channel can be obtained. Then the original feature $\textbf{x}_t$ is re-calibrated through channel element-wise product operation for HR spectral attention feature $\textbf{x}_{spe}$, defined as:
\setlength{\abovedisplayskip}{4pt}
\setlength{\belowdisplayskip}{4pt}
\begin{align}
\textbf{x}_{spe}  = \textbf{x}_t \odot {F}_{conv}^{1\times 1}(Sigmoid[f_c^{v}]).
\end{align}

\noindent\textbf{HR Spatial Attention.} For the input $\textbf{x}_t\in \mathbb{R}^{\it H \times \it W \times \it C}$, we obtain the \textit{key} vector defined as$f_s^{k}$ with the half channel resolution and the full spatial resolution. The \textit{query} vector defined as $f_s^{q}$ is regarded as the remapping factor to adjust the spatial attention for the \textit{value} vector $f_s^{v}$. Even if the global average pooling (GAP) sacrifices the channel resolution of $f_s^{q}$, the full spatial resolution of $f_s^{v}$ will bring HR features in the spatial dimension. These operations are defined as:
\setlength{\abovedisplayskip}{3pt}
\setlength{\belowdisplayskip}{3pt}
\begin{equation}
 \begin{split}
 f_s^{q} &= {F}_{GAP}({F}_{conv}^{1\times 1}(\textbf{x}_t))\in \mathbb{R}^{\textit{1} \times \textit{1} \times {\textit{C / }}2},\\
f_s^{k} &=  {F}_{conv}^{1\times 1}(\textbf{x}_t)\in \mathbb{R}^{\textit{C / }2 \times \it H \times \it W},\\
f_s^{v} &= Softmax[{F}_R(f_s^{q})]\otimes{F}_R(f_s^{k}) \in \mathbb{R}^{\it  1 \times \it HW},
 \end{split}
\end{equation}
where ${F}_{GAP}(\cdot)$ is the GAP function. As shown in Fig.~\ref{overall}, we re-calibrate the original feature $\textbf{x}_t$ by the weight factor of each spatial feature coordinate, and these weight factors come from the \textit{Sigmoid} activation value of $f_s^{v}$. Then the HR spatial attention feature $\textbf{x}_{spa}$ is calculated as:
\setlength{\abovedisplayskip}{3pt}
\setlength{\belowdisplayskip}{3pt}
\begin{align}
\textbf{x}_{spa}  = \textbf{x}_t \odot Sigmoid[{F}_R(f_s^{v})].
\end{align}

\noindent\textbf{Efficient Feature Fusion.} To further improve the feature utilization and interactivity within the spectral-spatial attention learning, we use an efficient fusion manner to group and re-interact the input features. Firstly, we fuse the spectral attention feature $\textbf{x}_{spe}$ and spatial attention feature $\textbf{x}_{spa}$:
\begin{align}
\textbf{x}_{f}  = \textbf{x}_{spe} + \textbf{x}_{spa},
\end{align}
where $\textbf{x}_{f}\in \mathbb{R}^{\rm H \times W \times C}$. Then, considering the diverse importance of different channels, we divide feature $\textbf{x}_{f}$ into $m$ groups, so the $\textbf{x}_{f}$ can be expressed as $[ \textbf{x}_f^{1}, \textbf{x}_f^{2}, \textbf{x}_f^{3}, ... , \textbf{x}_f^{m}]$. The channel number of each group $\textbf{x}^i_{f}$ ($i\in [1,m]$) is $C/m$.

As shown in Fig.~\ref{overall}, we replace the standard convolution with the depthwise-separable  convolution (DSC)~\cite{sandler2018mobilenetv2,howard2017mobilenets,chollet2017xception} to reduce the computational cost. For each group of features $\textbf{x}^i_{f}$, the salient features are extracted independently. After activating through the \textit{Softmax} layer, the corresponding weighting factor ${f}^i_{e}$ of each group can be expressed as:
\setlength{\abovedisplayskip}{3pt}
\setlength{\belowdisplayskip}{3pt}
\begin{equation}
 \begin{split}
{f}^i_{e} &= {F}_{DSC}(\textbf{x}^i_{f}) \\
 &= Softmax[{F}_{conv}^{PWC}(F_{MP}({F}_{conv}^{DWC}(\textbf{x}^i_{f})))],
\end{split}
\end{equation}
where ${F}_{conv}^{PWC}(\cdot)$ is the point-wise convolution (PWC), and ${F}_{conv}^{DWC}(\cdot)$ represents the depth-wise convolution (DWC). The $F_{MP}(\cdot)$ represents the max pooling function with a kernel size of $3 \times 3$. The normalized weight ${f}^i_{e}$ re-calibrates $\textbf{x}^i_{f}$. Then we introduce residual skip connection to further promote information flow. and get the re-interaction feature as:
\begin{align}
\Hat{\textbf{x}}^i_{f}  = {f}^i_{e}\textbf{x}^i_{f} + \textbf{x}^i_{f}.
\end{align}
We traverse each group and connect the feature maps of each group to obtain the final fusion feature $\Hat{\textbf{x}}_{f}$ as follows:
\setlength{\abovedisplayskip}{3pt}
\setlength{\belowdisplayskip}{3pt}
\begin{align}
\Hat{\textbf{x}}_{f}  = [\Hat{\textbf{x}}^1_{f}, \Hat{\textbf{x}}^2_{f}, \Hat{\textbf{x}}^3_{f}, ... , \Hat{\textbf{x}}^m_{f}],
\end{align}
where $[\cdot]$ denotes the concatenating operation. The efficient grouped DSC adjusts feature interactions of each group dynamically instead of equal treatment, which further ensures the extraction of high-resolution features. Efficient calculation greatly reduces parameter cost and calculation burden.
\vspace{-5mm}
\subsection{Frequency Domain Learning}
\vspace{-1mm}
The inherent bias of CNN makes it challenging to synthesize high-frequency features in SDL, which leads to the frequency domain discrepancy in other methods in Fig.\ref{fig:intro}. so we introduce dynamic FDL for frequency-level supervision.

\noindent\textbf{Discrete Fourier Transform.} DFT transforms the discrete signal from the time domain to the frequency domain to analyze the frequency structure. For a finite-length discrete 1D signal, the sine wave components of each frequency are obtained through the following correspondence:
\setlength{\abovedisplayskip}{3pt}
\setlength{\belowdisplayskip}{3pt}
\begin{equation}
F(w) = \textstyle \frac{1}{N} \sum_{n=0}^{N-1}f(n)e^{-j2 \pi {\frac{wn}{N}}},
\end{equation}
where $F(w)$ represent the frequency domain signal corresponding to the 1D discrete time domain signal $f(n)$.

\noindent\textbf{HSI Frequency Spectra Analysis.} We use the 2D DFT to convert the HSI to the frequency domain to reconstruct more high-frequency details. We define the ground truth and reconstructed HSI as $y_{gt}$ and $y_{pred}$ with the dimensions of $\mathbb{R}^{\it H \times W \times C}$. We calculate the frequency spectrum for each channel. In a specific channel $k$, the conversion relationship between spatial coordinates $(h,w,k)$ and frequency domain coordinates $(u,v)$ is expressed as:
\setlength{\abovedisplayskip}{2pt}
\setlength{\belowdisplayskip}{2pt}
\begin{equation}
 \begin{split}
\textbf{F}^k_{gt}(u,v) &=\textstyle {\sum_{h=0}^{H-1}}{\sum_{w=0}^{W-1}}y_{gt}(h,w,k)e^{-j2 \pi ({\frac{uh}{H}}+{\frac{vw}{W}})}, \\
\textbf{F}^k_{pred}(u,v) &=\textstyle {\sum_{h=0}^{H-1}}{\sum_{w=0}^{W-1}}y_{pred}(h,w,k)e^{-j2 \pi ({\frac{uh}{H}}+{\frac{vw}{W}})},\\
 \end{split}
\end{equation}
where the $\textbf{F}_{gt}$ and $\textbf{F}_{pred}$ are the frequency spectra of all channels corresponding to $y_{gt}$ and $y_{pred}$. As shown in Fig.~\ref{overall}, their frequency spectra visualization represents the severity of the grayscale changes. The structural textures and edges are mapped as high-frequency signals while the background as low-frequency signals. Therefore, we can easily manipulate the high-frequency or low-frequency information of the HSI. Then we introduce dynamic weights to make the network treat different frequencies adaptively.
\begin{table*}[htp!]
	\centering{
		\resizebox{\textwidth}{!}{
			\begin{tabular}{c|c|c|c|c|c|c|c|c|c}
				\hline
				\hline
				Method  & TwIST~\cite{twist} & GAP-TV~\cite{gap_tv}  & DeSCI~\cite{desci}  & $\lambda$-Net~\cite{lambda} & HSSP~\cite{hssp}  & DNU~\cite{dnu} & TSA-Net~\cite{tsa_net}  & DGSMP~\cite{gsm} &\bf HDNet (Ours)                  \\ \hline
				Scene1  & 25.16, 0.6996 & 26.82, 0.7544 & 27.13, 0.7479 & 30.10, 0.8492              & 31.48, 0.8577 & 31.72, 0.8634 & 32.03, 0.8920          & {33.26, 0.9152} &\bf 34.95, 0.9478 \\ \hline
				Scene2  & 23.02, 0.6038 & 22.89, 0.6103 & 23.04, 0.6198 & 28.49, 0.8054              & 31.09, 0.8422 & 31.13, 0.8464 & 31.00, 0.8583          & {32.09, 0.8977} &\bf32.52, 0.9531 \\ \hline
				Scene3  & 21.40, 0.7105 & 26.31, 0.8024 & 26.62, 0.8182 & 27.73, 0.8696              & 28.96, 0.8231 & 29.99, 0.8447 & 32.25, 0.9145          & {33.06, 0.9251} &\bf 34.52, 0.9569 \\ \hline
				Scene4  & 30.19, 0.8508 & 30.65, 0.8522 & 34.96, 0.8966 & 37.01, 0.9338              & 34.56, 0.9018 & 35.34, 0.9084 & 39.19, 0.9528          & {40.54, 0.9636} &\bf 43.00, 0.9810 \\ \hline
				Scene5  & 21.41, 0.6351 & 23.64, 0.7033 & 23.94, 0.7057 & 26.19, 0.8166              & 28.53, 0.8084 & 29.03, 0.8326 & {29.39, 0.8835} & 28.86, 0.8820 &\bf 32.49, 0.9565         \\ \hline
				Scene6  & 20.95, 0.6435 & 21.85, 0.6625 & 22.38, 0.6834 & 28.64, 0.8527              & 30.83, 0.8766 & 30.87, 0.8868 & 31.44, 0.9076          & {33.08, 0.9372} &\bf 35.96, 0.9645 \\ \hline
				Scene7  & 22.20, 0.6427 & 23.76, 0.6881 & 24.45, 0.7433 & 26.47, 0.8062              & 28.71, 0.8236 & 28.99, 0.8386 & 30.32, 0.8782          & \textbf{30.74}, 0.8860 &29.18, \bf 0.9373\\ \hline
				Scene8  & 21.82, 0.6495 & 21.98, 0.6547 & 22.03, 0.6725 & 26.09, 0.8307              & 30.09, 0.8811 & 30.13, 0.8845 & 29.35, 0.8884          & {31.55, 0.9234} &\bf 34.00, 0.9609 \\ \hline
				Scene9  & 22.42, 0.6902 & 22.63, 0.6815 & 24.56, 0.7320 & 27.50, 0.8258              & 30.43, 0.8676 & 31.03, 0.8760 & 30.01, 0.8901          & {31.66, 0.9110} &\bf 34.56, 0.9576\\ \hline
				Scene10 & 22.67, 0.5687 & 23.10, 0.5839 & 23.59, 0.5874 & 27.13, 0.8163              & 28.78, 0.8416 & 29.14, 0.8494 & 29.59, 0.8740          & {31.44, 0.9247} &\bf 32.22, 0.9500 \\ \hline
				Average & 23.12, 0.6694 & 24.36, 0.6993 & 25.27, 0.7207 & 28.53, 0.8406              & 30.35, 0.8524 & 30.74, 0.8631 & 31.46, 0.8939          & {32.63, 0.9166} &\bf 34.34, 0.9572\\ \hline\hline
	\end{tabular}}}
	\vspace{-3mm}
	\caption{The PSNR in dB (left entry in each cell) and SSIM (right entry in each cell) results of the test methods on 10 scenes.}
	\label{tab:simulation}
	\vspace{-4.5mm}
\end{table*}
\begin{figure*}[t]
	\begin{center}
		\begin{tabular}[t]{c} \hspace{-3.8mm}
			\includegraphics[width=1\textwidth]{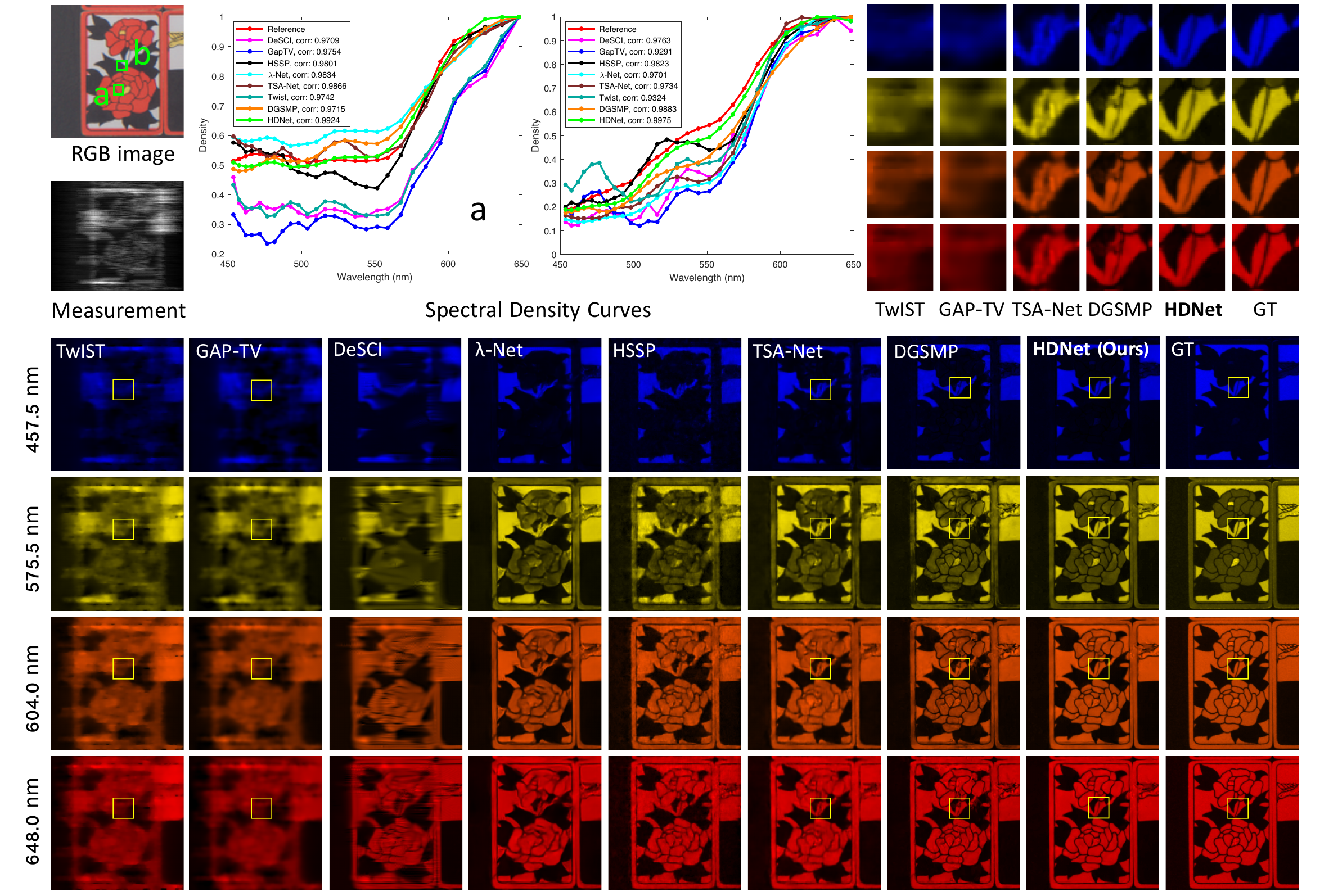}
		\end{tabular}
	\end{center}
	\vspace*{-8mm}
	\caption{\small Simulated HSI reconstruction comparisons of  \emph{Scene} 7 with 4 (out of 28) spectral channels. We show the spectral curves (top-medium) corresponding to the selected green boxes of the RGB image. Our HDNet reconstructs more visually pleasant detailed contents.}
	\label{fig:simulation}
	\vspace{-4mm}
\end{figure*}

\noindent\textbf{Frequency Distance Optimization.} We use a frequency distance coefficient $\alpha$ to make the distance correlation adjustable. In each channel $k$, the frequency distance between ground truth and predicted HSI is equivalent to the power distance between their spectrum, which is defined as:
\setlength{\abovedisplayskip}{3pt}
\setlength{\belowdisplayskip}{3pt}
\begin{align}
d^k(u,v) &= \textstyle{\|\textbf{F}^k_{gt}(u,v) -\textbf{F}^k_{pred}(u,v)\|}^\alpha.
\label{eq:distance}
\end{align}

The analysis of frequency-distance coefficients $\alpha$ is provided in Sec.~\ref{Ablation}. Then we define a dynamic weight factor $\theta(u,v)$ linearly related to the distance $d(u,v)$ to make the model pay more attention to the frequencies hard to be synthesized. Then the distance between the ground truth and the predicted HSI in a single channel $k$ is formulated as:
\setlength{\abovedisplayskip}{2pt}
\setlength{\belowdisplayskip}{2pt}
\begin{align}
d(\textbf{F}_{gt}^k,\textbf{F}_{pred}^k) &=\textstyle{\frac{1}{HW}}{\sum_{u=0}^{H-1}}{\sum_{v=0}^{W-1}}\theta^k(u,v)d^k(u,v),
\end{align}
where the $\theta^k(u,v)$ changes linearly with the absolute value of the $k$-th channel frequency distance $\sqrt{(|d^k(u,v)|)}$. We traverse $k=\{0, 1, 2, ..., C-1\}$ and sum each spectral distance to calculate the frequency domain loss in FDL as:
\setlength{\abovedisplayskip}{2pt}
\setlength{\belowdisplayskip}{2pt}
\begin{align}
L_{FDL}(\textbf{F}_{gt}, \textbf{F}_{pred}) &= \textstyle  {\sum_{k=0}^{C-1}}d(\textbf{F}_{gt}^k,\textbf{F}_{pred}^k).
\label{l_fdl}
\end{align}
\vspace{-5mm}
\subsection{Traning Objective}
\vspace{-2mm}
We choose the least absolute error as the loss in SDL, \textit{i.e.} ${L}_{SDL}(y_{gt}, y_{pred}) = \|y_{gt}-y_{pred}\|^1$. The loss in FDL is $L_{FDL}(\textbf{F}_{gt}, \textbf{F}_{pred})$ defined in Eq.~\eqref{l_fdl}. We introduce the weight factor $\lambda$ to balance SDL and FDL, and the total loss combined with the dual-domain learning is expressed as:
\setlength{\abovedisplayskip}{5pt}
\setlength{\belowdisplayskip}{5pt}
\begin{align}
L_{total}&= \textstyle L_{SDL}(y_{gt}, y_{pred}) + \lambda L_{FDL}(\textbf{F}_{gt}, \textbf{F}_{pred}).
\label{loss:total}
\end{align}
It is worth mentioning that how the model focuses on hard frequencies in FDL can be controlled by $\alpha$ in Eq.~\eqref{eq:distance}. The larger $\alpha$, the greater the penalty for the hard frequencies.
\vspace{-1mm}
\section{Experiments}
\vspace{-1mm}
\subsection{Experimental Setup}
\vspace{-2mm}
\noindent\textbf{Datasets.} We conduct experiments on two publicly available simulated HSI datasets CAVE~\cite{cave} and KAIST~\cite{kaist} for a fair comparison. The CAVE consists of 32 HSIs with 31 spectral bands with a spatial size of 512$\times$512, and the KAIST consists of 30 HSIs with 31 spectral channels at a size of 2704$\times$3376. Following TSA-Net~\cite{tsa_net} and DGSMP~\cite{gsm}, we use the same mask at a size of 256$\times$256 for simulation. 28 wavelengths ranging from 450nm to 650nm obtained by spectral interpolation manipulation are adopted. As with TSA-Net~\cite{tsa_net}, we use the CAVE dataset for training and select 10 scenes from KAIST for testing.

\noindent\textbf{Implementation Details.} We follow the same experimental settings as TSA-Net~\cite{tsa_net}. During the training, a patch at the size of 256$\times$256$\times$28 is randomly selected from the training 3D HSI datasets as labels. After mask modulation, the data cube is shifted in spatial with an accumulative two-pixel step and then summed up along the spectral dimension to generate the 2D measurement of size 256$\times$310. Random flipping and rotation are used for data argumentation. We use 32 RBs ($l=g=$16) and insert one SDL module in the middle. We set $\alpha=2$ in Eq.~\eqref{eq:distance} and $\lambda=0.7$ in Eq.~\eqref{loss:total}. The HDNet is optimized by ADAM~\cite{2014Adam} with the learning rate of $4 \times 10^{-4}$, which decreases linearly to half every 50 epochs. Our models are trained on NVIDIA GeForce RTX 2080 Ti GPU. Peak-signal-to-noise-ratio (PSNR) and structured similarity (SSIM)~\cite{wang2004image} are adopted as the metrics to evaluate the HSI reconstruction quantitatively. 
\vspace{-1mm}
\subsection{Comparison with Other Methods}
\vspace{-1mm}
\noindent\textbf{Quantitative Comparison:} We compare the HSI reconstruction of the proposed HDNet with 7 other SOTA methods, including three conventional methods (TwIST~\cite{twist}, GAP-TV~\cite{gap_tv}, and DeSCI~\cite{desci}) and four CNN-based methods ($\lambda$-Net~\cite{lambda}, HSSP~\cite{hssp}, DNU~\cite{dnu}, TSA-Net~\cite{tsa_net}, and DGSMP~\cite{gsm}). The quantitative results on 10 scenes of the KAIST dataset in terms of PSNR and SSIM are reported in Tab.~\ref{tab:simulation}. We can see that our HDNet significantly outperforms other methods. Specifically, our method surpasses the recent best competitor DGSMP by 1.71 dB in average PSNR and 0.0406 in average SSIM. When compared with the two deep unfolding algorithms HSSP and DNU, our HDNet is 3.99 dB and 3.60 dB higher, respectively. When compared with the two model-based methods TwIST and DeSCI, our HDNet achieves 11.22 dB and 9.98 dB performance gain.
Noted that although our HDNet's PSNR is slightly lower than DGSMP in Scene7, SSIM exceeds it by a large margin, which proves that the frequency domain optimization strategy we use is more focused on the refinement of perceptual quality and structural similarity. The complementary spatial-spectral domain and frequency domain further improve the reconstruction performance.  

\noindent\textbf{Visual Comparison:} We show the simulated HSI reconstruction comparisons of \emph{Scene} 7 with 4 (out of 28) spectral channels in Fig.~\ref{fig:simulation}. The density-wavelength spectral curves correspond to the green boxes identified as $a$ and $b$ of the RGB image. We calculate the curve correlation between all comparison methods and the reference truth. These quantitative results show that our reconstructed HSI is the closest and highest correlation to the ground truth (GT). Besides, we visualize the entire HSI and enlarge the selected yellow boxes in the upper right of Fig.~\ref{fig:simulation}. Our HDNet generates more visually pleasant results than previous methods, especially in the reconstruction of high-frequency structural content and spectral-dimension consistency, which benefit from pixel-level and frequency-level dual-domain learning.

\subsection{Ablation Study}
\label{Ablation}
\vspace{-1mm}
\noindent \textbf{Model Analysis.} There are some existing attention networks used for HSI reconstruction. We reports their parameters, spatial resolution, model complexity, and performance in Tab.~\ref{attentional-mechanism}. Noted that we retrain attentions in $\lambda$-Net~\cite{lambda} and TSA-Net~\cite{tsa_net} with the same baseline and the loss in Eq.~(\ref{loss:total}) for a fair comparison. Although the $\lambda$-Net treats each channel equally, the non-local spatial mechanism makes its parameter amount as high as 62.64M. TSA-Net sacrifices part of the channel resolution in exchange for computation complexity, but the used spatial-spectral self-attention also has a higher parameter burden. Our HDNet achieves the best trade-off between model performance and parameters, and it also maintains the greatest fine details and resolutions in both channel resolution (CR) and spatial resolution (SR). The parameter of our HDNet is 2.37M, which is less than one-eighteenth of TSA-Net while maintaining the same model complexity. These results show the superiority of our proposed HR attention mechanism.

\begin{figure}[t]
\vspace{-4.5mm}
\centering
\hspace{-2.5mm}
\includegraphics[width=84.5mm]{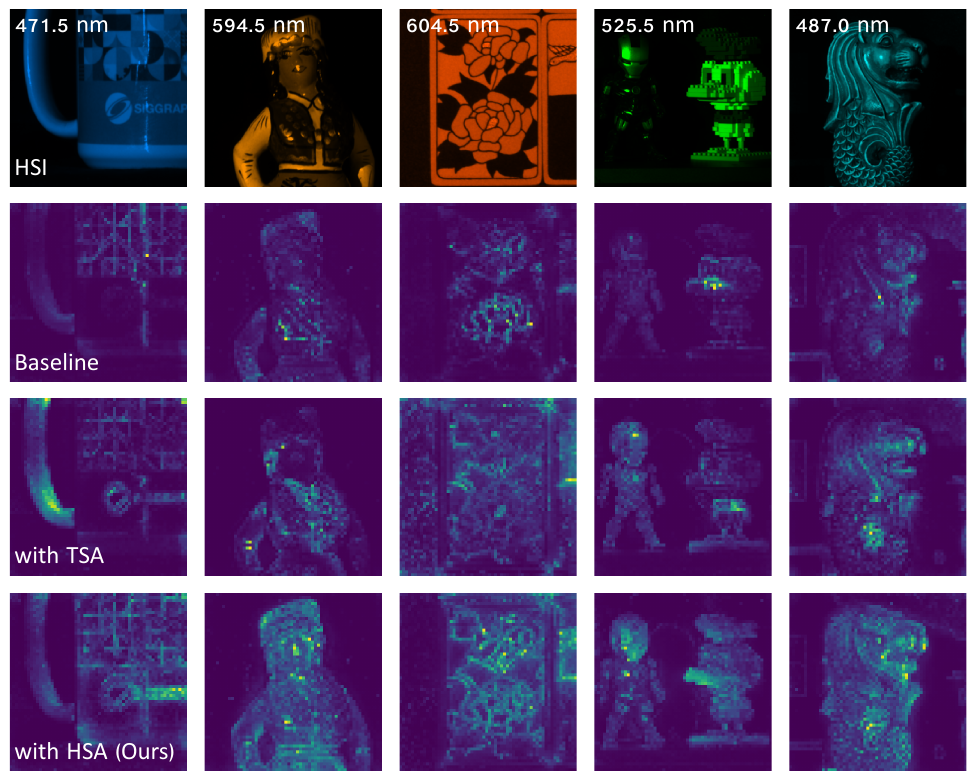}
\vspace{-8mm}
\caption{Feature visualization with different attention modules. }
\label{fig:att_vis}
\vspace{-5mm}
\end{figure}

\noindent\textbf{Attention Feature Visualization}: To show the advantages of our proposed HR spatial-spectral attention (HSA) in capturing HR fine-grained features more intuitively, we visualize the intermediate attention maps of different attention modules used for HSI reconstruction. We take ResNet~\cite{he2016deep} as the baseline, and then add the TSA and our HSA, respectively. The corresponding results are shown in Fig.~\ref{fig:att_vis}. Compared with the baseline, both TSA and HSA can enhance the extraction of salient features. However, TSA with lower resolution attention inevitably loses a lot of textures and edges and even focuses on the background by mistake. Our proposed HSA solves this problem well. The continuous HR attention in HSA allows the network to retain more high-frequency information and complete structure of HSI.

\vspace{0.5mm}
\noindent \textbf{Loss Weight Factor.} The weight factor $\lambda$ in Eq.~\eqref{loss:total} is introduced to adjust the importance of SDL and FDL dynamically. We analyze how the model performance changes with $\lambda$, and report the corresponding results in Tab.~\ref{loss-weight}. $\lambda=0$ means that the model only minimizes the spatial-spectral domain loss, and its unsatisfactory results indicate that frequency-level supervision is necessary. It can be seen that as the proportion of FDL loss increases, the model performance also increases. When $\lambda=0.7$, the model performance reaches the highest PSNR and SSIM performance. The performance degradation caused by the continued increase of $\lambda$ indicates that excessive constraints on the frequency will destroy the pixel-level optimization balance.
\begin{table}[t]
\vspace{-3mm}
	 \flushleft
	\resizebox{84mm}{!}{
		\begin{tabular}{l|cccccc}
			\hline
			Method & Params. & CR & SR & Complexity &  PSNR/SSIM\\
			\hline
			$\lambda$-Net~\cite{lambda} & 62.64M& 1 & [$H$,$W$]& $C^2WH + CW^2H^2 $ &30.85 / 0.9062 \\ 
			TSA-Net~\cite{tsa_net} & 44.25M & $C$ / 4 & [$H$,$W$]& $CWH$ & 32.68 / 0.9267\\ 
			HDNet (Ours) & \textbf{2.37M} & $C$ / 2 & [$H$,$W$]& $CWH$ & \textbf{34.34 / 0.9572} \\ 
			\hline 
		\end{tabular}
	}
    \vspace{-3.5mm}
   	\caption{Model analysis of different attention networks with different channel resolution (CR) and spatial resolution (SR).}
   	\label{attentional-mechanism}
\end{table}

\begin{figure}[t]
\vspace{-4mm}
 \flushleft
\hspace{-2.3mm}
\includegraphics[width=85mm]{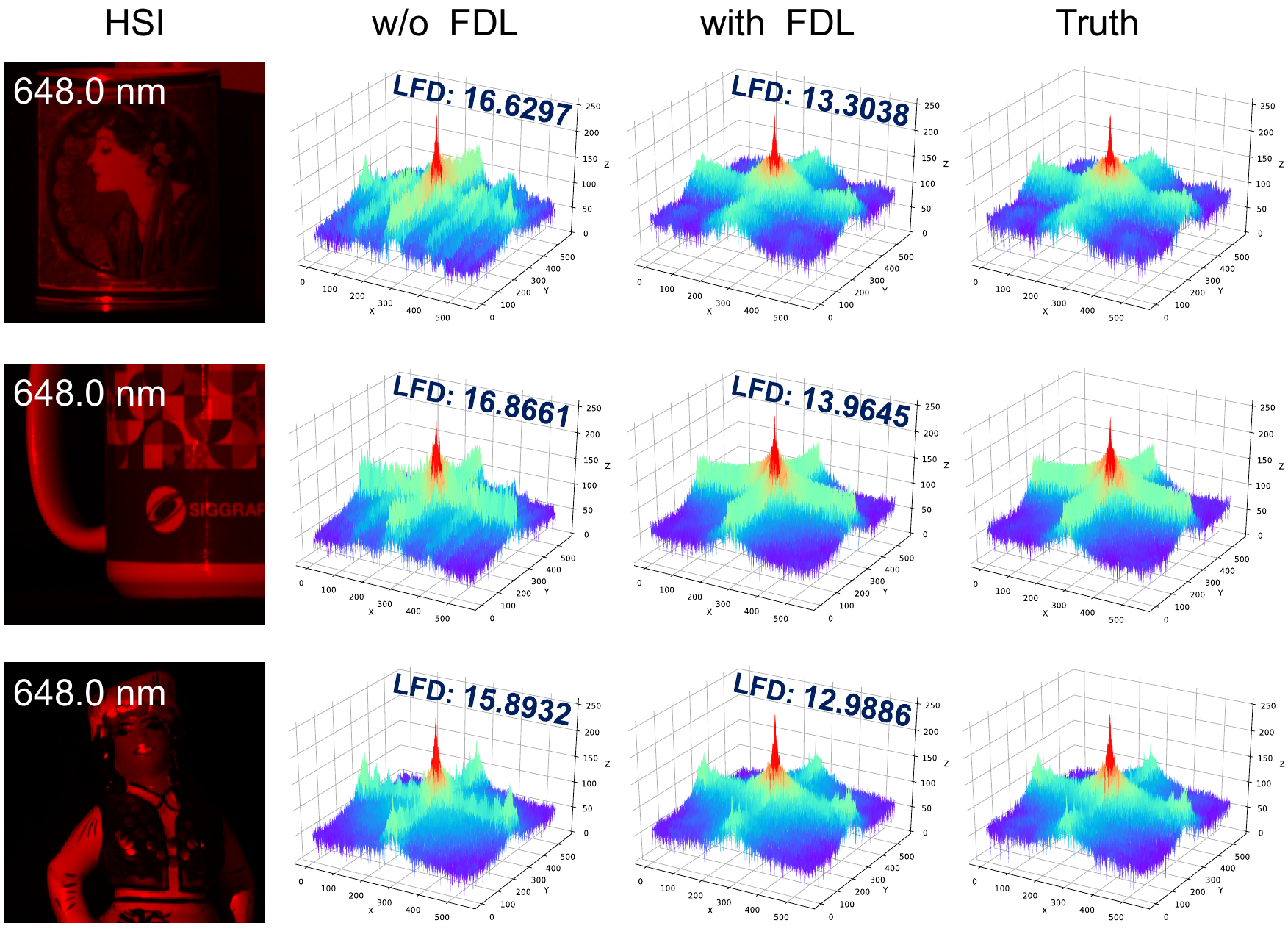}
\vspace{-7mm}
\caption{Frequency spectrum visualization with or without (w/o) FDL. The metric LFD is used to measure the frequency similarity. }
\label{fig:dfl}
\vspace{-5mm}
\end{figure}

\vspace{0.5mm}
\noindent\textbf{FDL Loss Ablation.} We calculate the log frequency distance (LFD) as a frequency-level metric to evaluate the spectrum difference between the reconstructed HSI and truth. The LFD has a logarithmic relationship with the frequency distance $d(u,v)$ in Eq.~\eqref{eq:distance}, which is calculated as:
\setlength{\abovedisplayskip}{4pt}
\setlength{\belowdisplayskip}{4pt}
\begin{align}
F_{LFD} = \textstyle log \Big[{\frac{1}{HW}}\Big({\sum_{u=0}^{H-1}}{\sum_{v=0}^{W-1}}{|d(u,v)|}\Big)+1 \Big].
\end{align}
As shown in Fig.~\ref{fig:dfl}, we visualize the 3D-spectra reconstructed with or without the FDL loss and provide the corresponding LFD. It can be seen that the reconstructed frequency 3D-spectra without frequency supervision has a ringing artifact, which will produce oscillations at the sharp brightness changes. Amplitude and phase distortion make different frequency components of HSI have different gain amplitude and relative displacement, which is manifested as deformed structure and color deviation in HSI. On the contrary, the 3D-spectra optimized with our proposed loss in FDL allow for more accurate frequency reconstruction and lower LFD, fitting the frequency statistics closer to truth. Fine-grained spectrum supervision further preserves more high-frequency information that is difficult to synthesize.
\begin{table}[t]
\vspace{-3.5mm}
	\centering
	\resizebox{84mm}{!}{
		\begin{tabular}{c|ccccccc}
			\hline
			$\lambda$ & ~0~ &~0.1~ & ~0.3~ & ~0.5~& ~0.7~ & ~0.9~ & ~1~ \\
			\hline
			SSIM $\uparrow$ & 0.9093& 0.9369 & 0.9498 & 0.9538 & \textbf{0.9572} & 0.9425 & 0.9399\\ 
			PSNR $\uparrow$ & 31.91&33.27 & 33.86 &34.05 &\textbf{34.34} & 33.75 & 33.52\\ 
			\hline 
		\end{tabular}
	}
    \vspace{-3mm}
   	\caption{Performance comparisons of different loss weight factor.}
   	\label{loss-weight}
   	\vspace{-0.5mm}
\end{table}
\begin{table}[t]
\vspace{-2.5mm}
	\centering
	\resizebox{84mm}{!}{
		\begin{tabular}{l|cccccc}
			\hline
			Metric & $\alpha$ = 0.1 & $\alpha$ = 0.3 & $\alpha$ = 0.5 & $\alpha$ = 1 & $\alpha$ = 2 & $\alpha$ = 3\\
			\hline
			LFD $\downarrow$ & 14.8633  &  14.3792 & 13.9825 &13.6571 &  \textbf{13.3238} &  15.0863   \\ 
			SSIM $\uparrow$& 0.9397& 0.9428 & 0.9543 & 0.9569 & \textbf{0.9572} & 0.9065\\ 
			PSNR $\uparrow$& 33.16& 33.51 & 34.14 &\textbf{34.40} & 34.34 &  31.89\\ 
			\hline 
		\end{tabular}
	}
    \vspace{-3mm}
   	\caption{Model performance comparison using different coefficients to calculate the spectrum distances in frequency domain.}
   	\label{distance-factor}
\end{table}

\begin{figure*}[t]
	\begin{center}
		\begin{tabular}[t]{c} \hspace{-3.8mm}
			\includegraphics[width=1\textwidth]{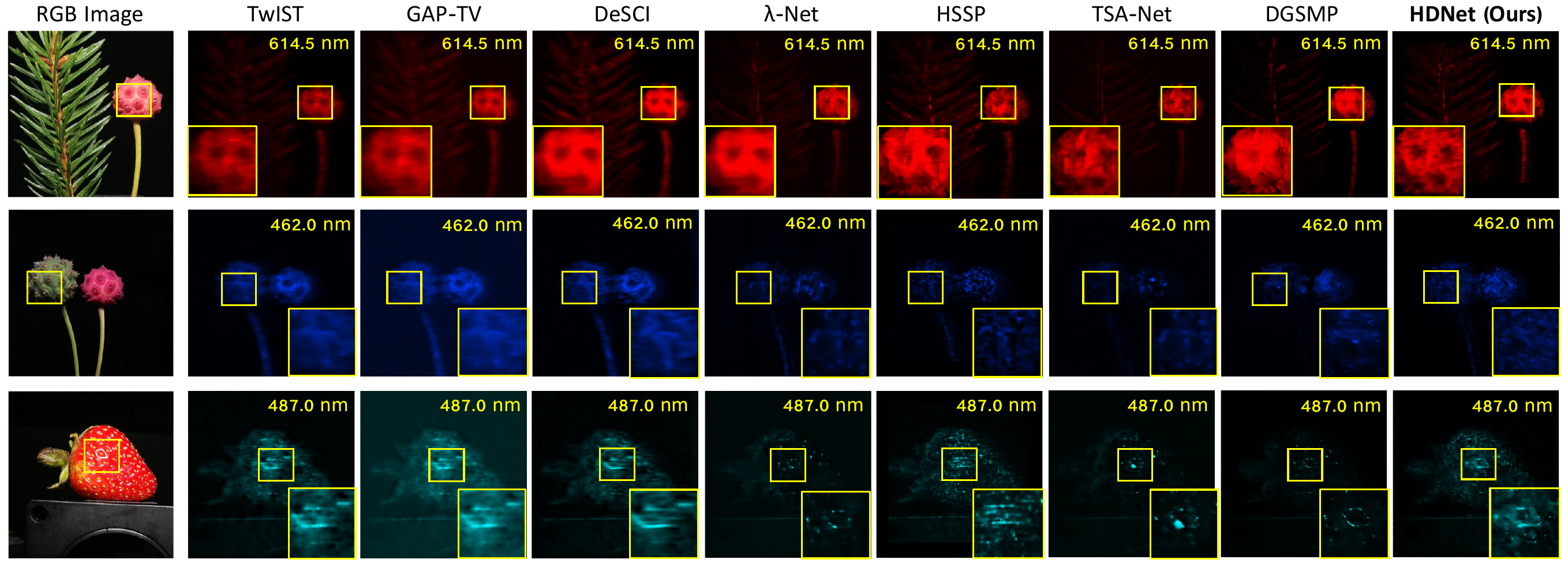}
		\end{tabular}
	\end{center}
	\vspace*{-8mm}
	\caption{\small Real HSI reconstruction comparison of a randomly selected channel from 3 scenes. HDNet restores more high-frequency details.}
	\label{fig:real}
	\vspace{-5mm}
\end{figure*}
\begin{figure}[t!]
\vspace{-1.5mm}
\flushleft
\includegraphics[width=85mm]{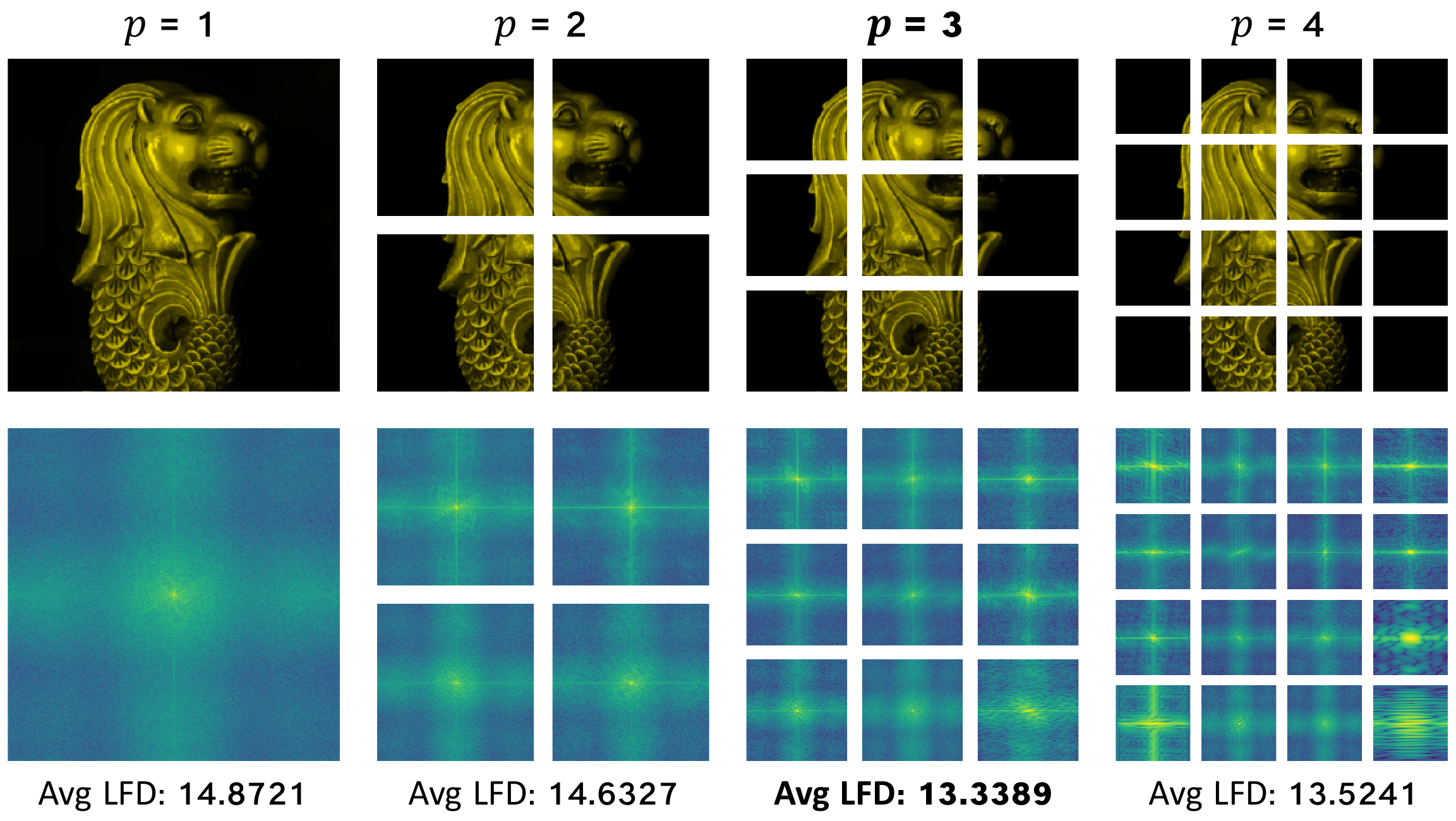}
\vspace{-7mm}
\caption{Patch-based spectrum visualization. $p$ is the patch number in each direction. We calculate the average LFD of all patches. }
\label{fig:patch_vis}
\vspace{-1mm}
\end{figure}

\noindent \textbf{Frequency Distance Coefficient.} The far frequency distance between reconstructed HSI and truth represents the inaccurate fit, so we introduce a frequency distance coefficient $\alpha$ in Eq.~\eqref{eq:distance} to control the model's focus on frequencies that have not been reconstructed well. The larger $\alpha$ is, the greater the model will penalize the underfitting frequency. We report the model performance corresponding to different coefficients in Tab.~\ref{distance-factor}. When $\alpha=1$, the model obtains the highest PSNR, and when $\alpha=2$, the model obtains the best SSIM and LFD performance. Smaller $\alpha$ results in weaker frequency penalty and slightly lower performance, but the larger $\alpha$ brings more stringent FDL supervision and excessive constraint, which will lead to HSI distortion and performance degradation. In order for the model to focus on both structural similarity and perceptual quality, we set $\alpha=2$ to balance the visual and quantitative results.

\begin{table}
\vspace{-4.5mm}
	\centering
	\resizebox{84mm}{!}{
		\begin{tabular}{l|cccccc}
			\hline
			Metric & $p$ = 1 & $p$ = 2 & $p$ = 3 & $p$ = 4 & $p$ = 5 & $p$ = 6 \\
			\hline
			LFD $\downarrow$& 14.7954  &  14.6287 &  \textbf{13.3238} &  13.5982 &  14.6391 &  14.9625 \\ 
			SSIM $\uparrow$& 0.9193& 0.9344 & \textbf{0.9572} & 0.9378 & 0.9304 & 0.9166\\ 
			PSNR $\uparrow$& 32.96&33.29 & \textbf{34.34} &33.69 &33.15 & 32.67\\ 
			\hline 
		\end{tabular}
	}
    \vspace{-3mm}
   	\caption{Model performance comparisons of different patch size.}
   	\label{Patch-based}
   	\vspace{-7mm}
\end{table}
\noindent \textbf{Patch-based Frequency Spectrum.} To further analyze the frequency characteristics of HSI, we replace the entire image spectrum calculation with patch-based calculation. The original HSI is cropped into $p \times p$ patches. The new frequency domain distance $\Hat{L}_{FDL}(F_{gt}, F_{pred})$ of each paired images will be redefined as the average value of each paired patches in the current channel. The Eq.~\eqref{l_fdl} is modified as:
\begin{align}
\Hat{L}_{FDL}(F_{gt}, F_{pred}) &=\textstyle {\sum_{k=0}^{C-1}}[\frac{1}{p^2}{\sum_{j=1}^{p^2}}d(F_{gt}^{kj},F_{pred}^{kj})].
\end{align}
We analyze the performance when $p=\{1,2,3,4,5,6\}$ in Tab.~\ref{Patch-based} and visualize part of the spectra in Fig.~\ref{fig:patch_vis}. The smaller patch-based subdivision brings a finer spectrum reconstruction, and the performance on narrowing the frequency domain gap shown by LFD also becomes better. However, the HSI reconstruction worsens when $p>3$. The visualization of $p=4$ in Fig.~\ref{fig:patch_vis} shows that a too small patch brings limited spectrum representation and biased supervision. So we set $p = 3$ to appropriately ameliorate the refinement of HSI.

\subsection{Real HSI Reconstrcution}
\vspace{-1mm}
We also apply the proposed HDNet for real HSI reconstruction. The dataset is collected by the real HSI system designed in TSA-Net~\cite{tsa_net}. Each HSI has 28 spectral channels with wavelengths ranging from 450nm to 650nm and has 54-pixel dispersion in the column dimension. The measurement used as input is at a spatial size of 660$\times$714. Following TSA-Net~\cite{tsa_net}, we re-train the HDNet with the real mask on the CAVE and KAIST datasets jointly. We also inject 11-bit shot noise on the 2D compressive image to simulate the real situations. Due to the lack of ground truth, we only compare the qualitative results of our HDNet with other methods. Reconstruction results of one channel randomly selected from 3 real scenes are shown in Fig.~\ref{fig:real}. Previous methods with only coarse spatial domain loss produce excessive smoothing and distortion of high-frequency details. The proposed HDNet generates more visually pleasant results by recovering more HR structures and high-frequency textures, which benefits from the pixel-level fine-grained and frequency-level refinement. Our robust results in the real dataset show good model generalization.
\vspace{-2mm}
\section{Conclusion}
\vspace{-1mm}
In this paper, we propose a high-resolution dual-domain learning network (HDNet) that includes spatial-spectral domain learning and frequency domain learning for HSI reconstruction from compressive measurements. The fine-grained pixel-level prediction is obtained by efficiently designing the HR spatial-spectral attention and feature fusion module. To solve the visual difference caused by the pixel-level loss, we introduce dynamically adjusted frequency-level supervision for the first time to narrow the frequency domain discrepancy between reconstructed HSI and the truth. HDNet exhausts the representation capacity within the dual-domain. Extensive visual analysis and quantitative experiments prove that the HDNet obtains superior results in both pixel-level and frequency-level HSI reconstruction.

{\small
\balance
\bibliographystyle{ieee_fullname}
\bibliography{reference}
}

\end{document}